\documentstyle[12pt,a4wide,epsfig,amssymb]{article}

\title{
\vspace*{-2.5cm}
\begin{flushright}
{\normalsize CPT-2000/P.4084\\[-2pt]
URCHEP-T292\\[-14pt]
}
\end{flushright}
\vspace{1.8cm}
{\Large \bf Supersymmetric Triplet Higgs Model of \\
Neutrino Masses and Leptogenesis\\}
\vspace*{0.8cm}
}
\author{Thomas Hambye$^{a,b\,}$, Ernest Ma$^{c\,}$, 
and Utpal Sarkar$^{d\,}$\\[0.5cm]
\normalsize $a$: {\it Centre de Physique Th\'eorique, 
CNRS Luminy, 13288 Marseille, France}\\[1mm] 
\normalsize $b$: {\it INFN - Laboratori Nazionali di Frascati,
00044 Frascati, Italy}\\[1mm] 
\normalsize $c$: {\it Department of Physics, University of California,
          Riverside, California 92521, USA}\\
\normalsize $d$: {\it Physical Research Laboratory, 
Ahmedabad 380 009, India}\\[2cm] 
}
\date{}
\begin{document}
\maketitle
\thispagestyle{empty}
\vspace*{4cm}
\begin{abstract}\
We construct a supersymmetric version of the triplet Higgs model for
neutrino masses, which can generate a baryon asymmetry of the Universe 
through lepton-number violation and is consistent with the gravitino 
constraints. 
\end{abstract}

\newpage
\baselineskip 24pt

\section{Introduction}

The first definite evidence for physics beyond the standard model came from 
the recent evidence for the mass of the neutrinos. The atmospheric neutrino 
anomaly \cite{atm}, as observed by the SuperKamiokande experiment, has 
established that there is a mass-squared difference between the muon 
neutrino and the tau neutrino.  On the other hand, the solar neutrino 
problem \cite{solar} implies a mass-squared difference between the electron 
neutrino and the other two active neutrinos.  Hence it has now been 
established that at least two neutrinos are massive. 

The mass-squared differences between the different generations of neutrinos
have to be very small, but the mixing angles large, to explain the 
atmospheric and solar neutrino anomalies.  The required masses for the 
neutrinos are several orders of magnitude smaller than those of other 
fermions, which are all Dirac particles.  The smallness of the neutrino 
mass is naturally explained if the neutrinos are Majorana particles 
\cite{ma}, hence lepton number is not conserved and that should be due to 
some physics beyond the standard model.  There are several motivations for 
lepton-number violation in Nature \cite{witten}.  In addition, the 
associated lepton-number violation may have the added virtue of accounting 
for the present observed baryon asymmetry of the Universe.  One model 
of neutrino mass having this virtue is the triplet 
Higgs model \cite{triplet}.  In the 
nonsupersymmetric case, this model has been studied in detail and found to 
share all the interesting features of other models of neutrino mass.  
Moreover, in theories with large extra dimensions \cite{extra}, this 
mechanism happens to be the only one which gives Majorana (rather than Dirac) 
masses to the neutrinos \cite{extrip}. In this article we will study the 
supersymmetric version of this model. 

In the supersymmetric version of the triplet Higgs model, there are several 
new aspects.  Similar to the requirement of two Higgs doublets in the 
supersymmetric extension of the standard model, we now have two Higgs 
triplets.  Only one of them couples to the leptons, but it can acquire
a vacuum expectation value ($vev$) only if the other Higgs triplet is present. 
This is related to the fact that a mass term in the superpotential requires 
two triplet Higgs superfields, which are of course also necessary for 
anomaly cancellation.  This mass term connecting the two triplet superfields 
in the superpotential also allows a trilinear coupling to exist between 
two scalar doublets and the scalar triplet which couples to leptons, which is 
necessary for neutrino mass as well as leptogenesis \cite{triplet}.  In the 
present supersymmetric version of the triplet Higgs model, we must consider 
the decays of both heavy triplets.  Note that supersymmetry is not yet 
broken at this energy scale. There are now also several new diagrams which
contribute to the CP violation. 

Another important feature of the supersymmetric model comes from the 
constraints of nucleosynthesis.  In supersymmetric models there is a 
strong bound on the scale of inflation from nucleosynthesis due to the 
gravitino problem \cite{gravitino}.  This means that baryogenesis has to 
occur at temperatures below about $10^{11}$ GeV.   On the other hand, in
the triplet Higgs model, the gauge interactions of the triplet Higgs 
scalars and fermions bring their number densities to equilibrium at 
temperatures below $\sim 10^{12}$ GeV.  This naive order-of-magnitude estimate 
thus implies that the supersymmetric triplet Higgs model of leptogenesis 
is probably not consistent with the gravitino constraints \cite{david}. 
However, detailed calculations give several possible ways out of this problem. 
In the following we will consider the cases where this potential problem is 
first ignored and then taken into account. We point out here that the 
supersymmetric triplet Higgs model can evade this problem of gravitinos 
when the masses of the triplet Higgs superfields are moderately degenerate.

In Section 2 we introduce the model and describe its consequences 
for neutrino masses. 
Then in Section 3 we calculate the
amount of CP violation in the decays of the triplet Higgs scalars and fermions 
which can generate a lepton asymmetry of the Universe. In Section 4 we solve 
the Boltzmann equations to calculate the evolution of the lepton asymmetry
and present our results. In Section 5 the gravitino problem is discussed.
Finally in Section 6 we summarize and conclude.

\section{The Model}

The Majorana masses of the neutrinos can be generated by extending
the standard model to include a triplet Higgs scalar, which acquires a
small $vev$ and couples to two leptons.  If lepton number was spontaneously 
broken by this $vev$ \cite{trip}, the so-called triplet Majoron (i.e. the 
resulting massless Goldstone boson) coupling to the Z boson 
would be predicted.  
This scenario is now ruled out by the known invisible $Z$ width
\cite{renton}.  Moreover, such models do not explain the present observed 
baryon asymmetry of the Universe.  A new scenario was then proposed in which 
lepton number is broken explicitly at a very high energy scale \cite{triplet}. 
The triplet Higgs scalar would then be extremely heavy.  However, it acquires 
a very tiny $vev$ through its lepton-number violating trilinear coupling 
to the standard-model Higgs doublet, which can then give a small
Majorana mass to the neutrinos. The decays of the triplet Higgs
scalars also generate a lepton asymmetry of the Universe, which
gets converted to a baryon asymmetry of the Universe before
the electroweak phase transition. 

To implement the triplet Higgs mechanism in a supersymmetric model, we need 
to extend the supersymmetric standard model to include two triplet Higgs 
superfields.  Since we want these fields to be very heavy, supersymmetry 
should be unbroken at that stage and the generation of the lepton asymmetry 
will not depend on the supersymmetry-breaking mechanism.  We also assume 
that R-parity is not violated, so that there is no other source of 
lepton-number violation except for the Yukawa couplings of the triplet 
Higgs superfields. 
We introduce one triplet $\hat \xi_1 ([\hat \xi_1^{++},\hat \xi_1^{+},\hat 
\xi_1^{0}] \equiv [1,3,1]$ under $SU(3)_c \times SU(2)_L \times U(1)_Y)$
and another 
triplet $\hat \xi_2 ( [\hat \xi_2^{0},\hat \xi_2^{-},\hat \xi_2^{--}] 
\equiv [1,3,-1])$ so that a mass term $M \hat{\xi}_1 \hat{\xi}_2$ may appear 
in the superpotential.
However, CP violation is not possible with just these two Higgs triplets. 
For that, we need two of each type of the above Higgs triplets.  So, if heavy 
triplet superfields are used to generate neutrino masses as well as a 
lepton asymmetry of the Universe, there should be at least four: 
$\hat \xi_1^a ([\hat \xi_1^{a++},\hat \xi_1^{a+},\hat \xi_1^{a0}] \equiv 
[1,3,1])$ and $\hat \xi_2^a ( [\hat \xi_2^{a0},\hat \xi_2^{a-},\hat 
\xi_2^{a--}] \equiv [1,3,-1])$, where $a=1,2$ corresponds to the two
scalar superfields, whose mixing gives CP violation for generating 
the lepton asymmetry of the Universe. 

The essential part of the superpotential for the interactions of these
scalar superfields with the lepton superfields $\hat L_i \equiv 
\pmatrix{\nu_{Li} & e^-_{Li}} \equiv [1,2,-1/2]$ and the standard Higgs 
doublets $\hat H_1 ( [\phi_1^0, \phi_1^-]
\equiv [1,2,-1/2])$ and $\hat H_2 (  [\phi_2^+, \phi_2^0]
\equiv [1,2,1/2])$ is given by
\begin{equation}
W=M_{ab} \hat \xi_1^a \hat \xi_2^b + f^a_{ij} \hat L_i \hat L_j \hat \xi_1^a + 
h_1^a \hat H_1 \hat H_1 \hat \xi_1^a
+ h_2^a \hat  H_2 \hat H_2 \hat \xi_2^a + \mu \hat H_1 \hat H_2 + ...
\label{superpot}
\end{equation}
where $i = 1,2,3$ is the generation index. The first term gives 
masses to the triplets. The condition for leptogenesis
and neutrino masses would determine this scale $M$.  The next
term gives the Yukawa couplings of the triplet Higgs scalar superfield
with the left-handed lepton chiral superfields of the three generations. 
When the scalars $\xi_1^a$ acquire vacuum expectation value ($vev$s),
this term gives Majorana masses to the neutrinos. The next two terms
give small $vev$s to the triplet Higgs scalars. 

The scalars $\xi_1^a$ couple to two leptons, to two
Higgsinos $\tilde H_1$, to two scalars $H_2$ and to a $H_1 H_2$ pair.
The scalars $\xi_2^a$ couple to two Higgsinos $\tilde H_2$, 
to two sleptons, to two scalars $H_1$ and to a $H_1 H_2$ pair.
This simultaneous decay of the triplets to products with
different lepton numbers breaks lepton number explicitly. Thus
the scale of lepton-number violation is the same as the mass of the
triplet Higgs scalars, which is very heavy, say of the order 
$\sim O(10^{9}-10^{14})$ GeV. However, since $SU(2)_L$ is unbroken at 
this scale, these fields do not acquire any $vev$. Only after
the electroweak symmetry breaking is there an induced tiny 
$vev$ for these scalars and the neutrinos would acquire mass.

The $vev$s of the triplet Higgs scalars are obtained from the 
vanishing of the $F-$terms, which corresponds to the minima of the 
potential. From the conditions $F_{\xi_1^a} = F_{\xi_2^a} = 0$,
and assuming that R-parity is conserved (so that 
the sneutrinos do not acquire any $vev$), we get
\begin{eqnarray}
&& F_{\xi_1^a} = M_{ab} \xi_2^b + f^a_{ij} \tilde L_i \tilde L_j
+ h_1^a H_1 H_1 = 0 \Longrightarrow
\langle \xi_2^b \rangle = u_2^b = - M^{-1}_{ba} h_1^a \langle H_1 \rangle^2  
= - M^{-1}_{ba} h_1^a v_1^2, \nonumber \\
&& F_{\xi_2^a} = M_{ab} \xi_1^b 
+ h_2^a H_2 H_2 = 0 \Longrightarrow  
\langle \xi_1^b \rangle = u_1^b = - M^{-1}_{ba} h_2^a \langle H_2 \rangle^2 
 = - M^{-1}_{ba} h_2^a v_2^2  \label{trvev} .
\end{eqnarray}
Since the masses of the triplet scalar fields are several orders 
of magnitude higher than the electroweak symmetry breaking scale
$v$, the effective $vev$ of the triplet Higgs fields are several orders
of magnitude smaller than $v = 246$ GeV.\footnote{The smallness of these $vev$s
makes this triplet model perfectly consistent with the usual
constraints on additional triplets coming from the measurement of the $\rho$ 
parameter at LEP.} Since these $vev$s give masses to
the neutrinos, the smallness of the neutrino mass is now directly
related to the large lepton-number violating scale.

The $vev$s of the triplet scalars will give a mass to the 
neutrinos given by
\begin{equation}
(m_\nu)_{ij} = \sum_a 2 f^a_{ij} u_1^a = 
\sum_{a,b}- 2 f^a_{ij} M^{-1}_{ab} h_2^b v_2^2.
\label{numass}
\end{equation}
Since the leptons do not couple with the other triplet scalar
$\xi_2$, there is no contribution to the neutrino mass from $u_2^a$.
Since the lepton number is now broken at a very large scale
explicitly, there is no Majoron in this scenario. There is one
would-be Majoron, which becomes too heavy to affect any low-energy
phenomenology. This makes it consistent with the measured invisible 
$Z$ width from LEP (Large Electron Positron Collider) at CERN. 

\begin{figure}[t]
\centerline{
\epsfxsize = 0.8\textwidth \epsffile{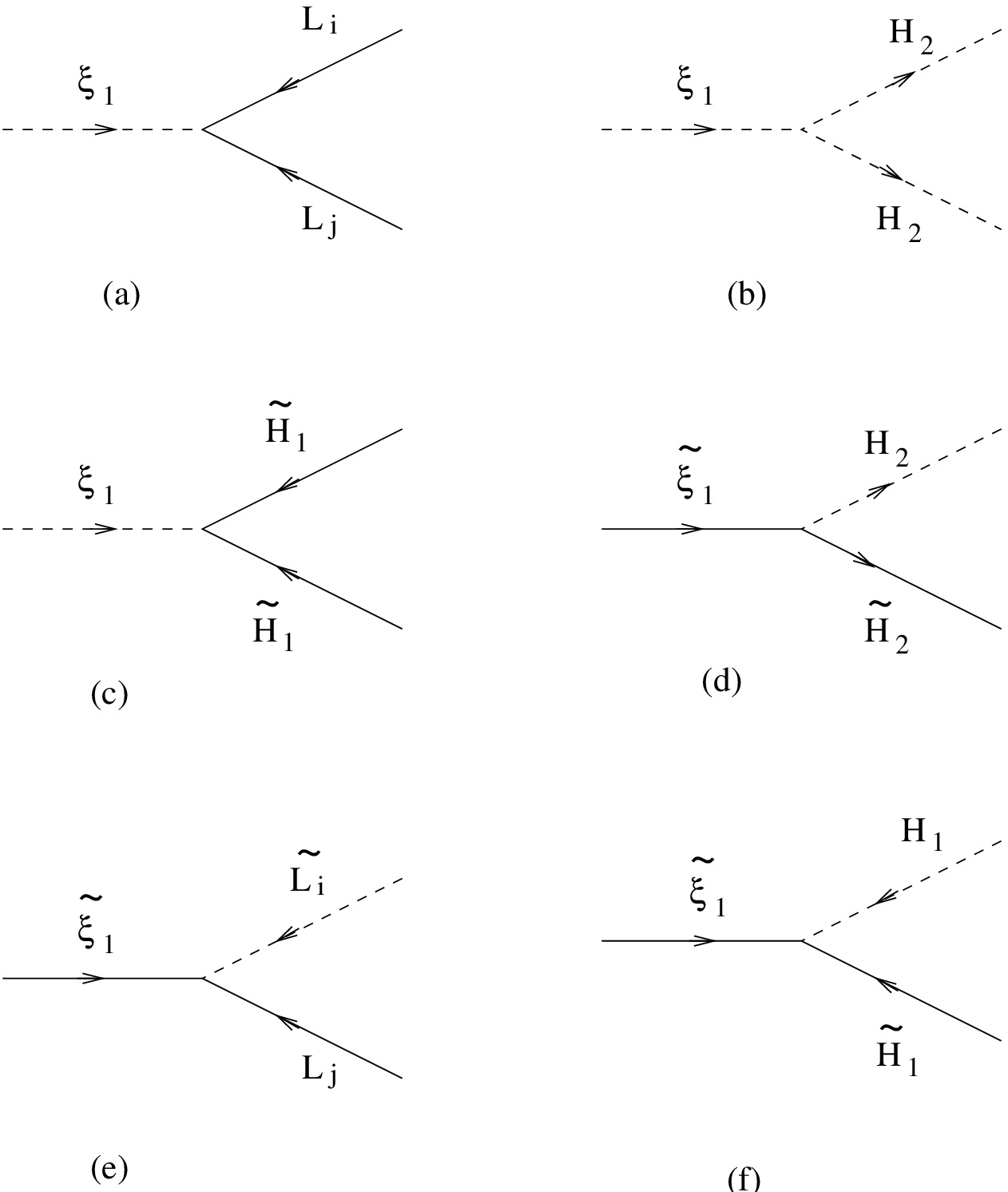}
}
\caption{ Tree level diagrams for the decay of $\hat \xi_1$. }
\label{strfg1}
\end{figure}

\begin{figure}[t]
\centerline{
\epsfxsize = 0.8\textwidth \epsffile{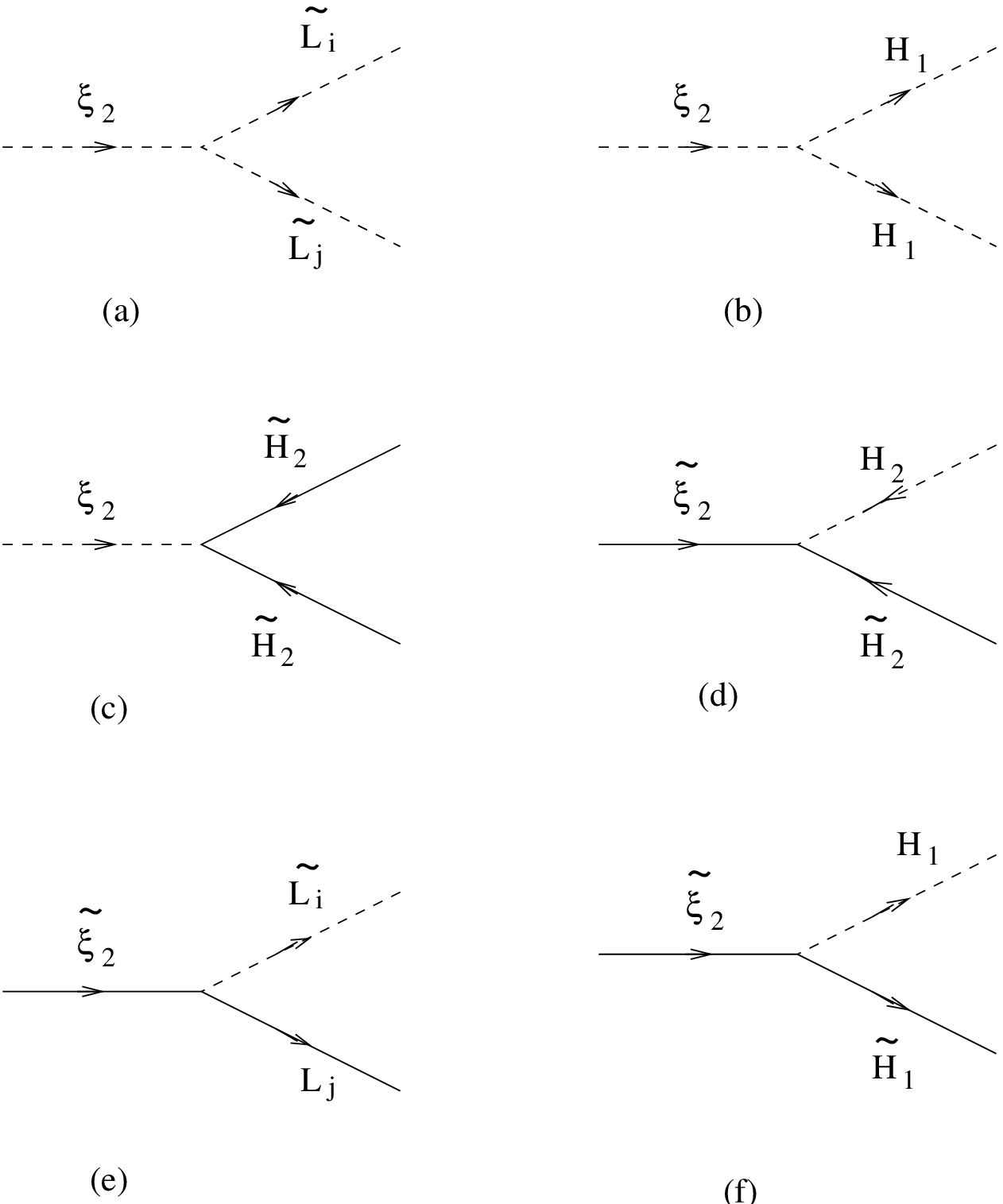}
}
\caption{ Tree level diagrams for the decay of $\hat \xi_2$.}
\label{strfg3}
\end{figure}

The decay of these scalars to two leptons or two Higgsinos can be
read off from the $F-$terms in the superpotential.  The decays 
of these scalars into two sleptons and the standard-model Higgs doublets can 
be read off from the relevant part of the scalar potential, 
\begin{eqnarray}
V &=& | M_{ab} \xi_2^b + f^a_{ij} \tilde L_i \tilde L_j 
+ h_1^a H_1 H_1|^2 + |M_{ab} \xi_1^b + h_2^a H_2 H_2 |^2
\nonumber \\ &&
+ |2 h_1^a H_1 \xi_1^a + \mu H_2 + ... |^2
+ | 2 h_2^a H_2 \xi_2^a + \mu H_1 + ... |^2 + ...
\label{scalpot}
\end{eqnarray}
The various decay modes of the scalar and fermionic components of
the triplet scalar superfields are listed below and shown in
Figures \ref{strfg1} and \ref{strfg3}. The decay modes of the $\hat \xi_1^a$
(i.e. the scalars $\xi_1^{a++}$ and the 
fermions $\tilde \xi_1^{a++}$) are
\begin{equation}
{\xi_1^a}^{++} \rightarrow \left\{ \begin{array} {l@{\quad}l} L_i^+ L_j^+ & 
(L = -2) \\ H_2^+ H_2^+ & (L= 0) \\ \tilde H_1^+ \tilde H_1^+ & (L=0) 
\end{array} \right.
\end{equation}
and 
\begin{equation}
\tilde {\xi_1^a}^{++} \rightarrow \left\{ \begin{array} {l@{\quad}l} 
\tilde L_i^+ L_j^+ & (L = -2) \\ \tilde H_2^+ H_2^+ & (L= 0) \\ 
H_1^+ \tilde H_1^+ & (L=0) \end{array} \right. ,
\end{equation}
while the decay modes of $\hat \xi_2^a$
are
\begin{equation}
{\xi_2^a}^{++} \rightarrow \left\{ \begin{array} {l@{\quad}l} 
\tilde L_i^+ \tilde L_j^+ & 
(L = -2) \\ \tilde H_2^+ \tilde H_2^+ & (L= 0) \\ H_1^+ H_1^+ & (L=0) 
\end{array} \right.
\end{equation}
and 
\begin{equation}
\tilde {\xi_2^a}^{++} \rightarrow \left\{ \begin{array} {l@{\quad}l} 
\tilde L_i^+ L_j^+ & (L = -2) \\ \tilde H_2^+ H_2^+ & (L= 0) \\ 
H_1^+ \tilde H_1^+ & (L=0) \end{array} \right. .
\end{equation}
The couplings entering in the various decay modes can be read off
from the superpotential. Note that we don't consider the decays
proportional to $\mu^2$ (i.e. to a $\hat H_1 \hat H_2$ pair) which 
are negligible. If there is CP violation and the decays
satisfy the out-of-equilibrium condition, then these decays can
generate a lepton asymmetry of the Universe \cite{sakh,kolb}. This
lepton asymmetry can then get converted to a baryon asymmetry of 
the Universe \cite{fy}.

When this lepton asymmetry is generated, the $B+L$ violating (but $B-L$ 
conserving) sphaleron transitions are taking place at a very fast rate 
\cite{krs}.  In fact, during the period
$$ 10^{12} ~{\rm GeV} ~ > T > ~ 10^2 ~ {\rm GeV} $$
the anomalous $B+L$ violating sphaleron processes remain in equilibrium.
During this period,  any lepton asymmetry of the Universe would
be equivalent to the $B-L$ asymmetry. The sphaleron interactions would then
convert this lepton asymmetry to a baryon asymmetry of the Universe 
within this period \cite{ht}. 

\section{CP Asymmetry in Triplet Higgs Decay}

The various decay modes
of $\xi_{1}^a$ and $\tilde \xi_{1}^a$ are given
in Figure \ref{strfg1} and  the decay modes of $\xi_{2}^a$ and 
$\tilde \xi_{2}^a$ are given in Figure \ref{strfg3}. 
The simultaneous decay of the triplet Higgs scalars
or the triplet Higgsinos into states with lepton
number 0 (two scalar Higgs doublets or Higgsinos) and with lepton number 2 
(two leptons or sleptons) implies lepton-number violation. 
For CP violation, the tree-level diagrams by themselves are not enough.
Even if the couplings are complex, the probability will be positive
definite and hence there will not be any CP violation.  However,
if there are one-loop diagrams, which interfere with these tree-level 
diagrams, then the interference may be complex, which gives
the CP violation. 

\begin{figure}[t]
\centerline{
\epsfxsize = 0.7\textwidth \epsffile{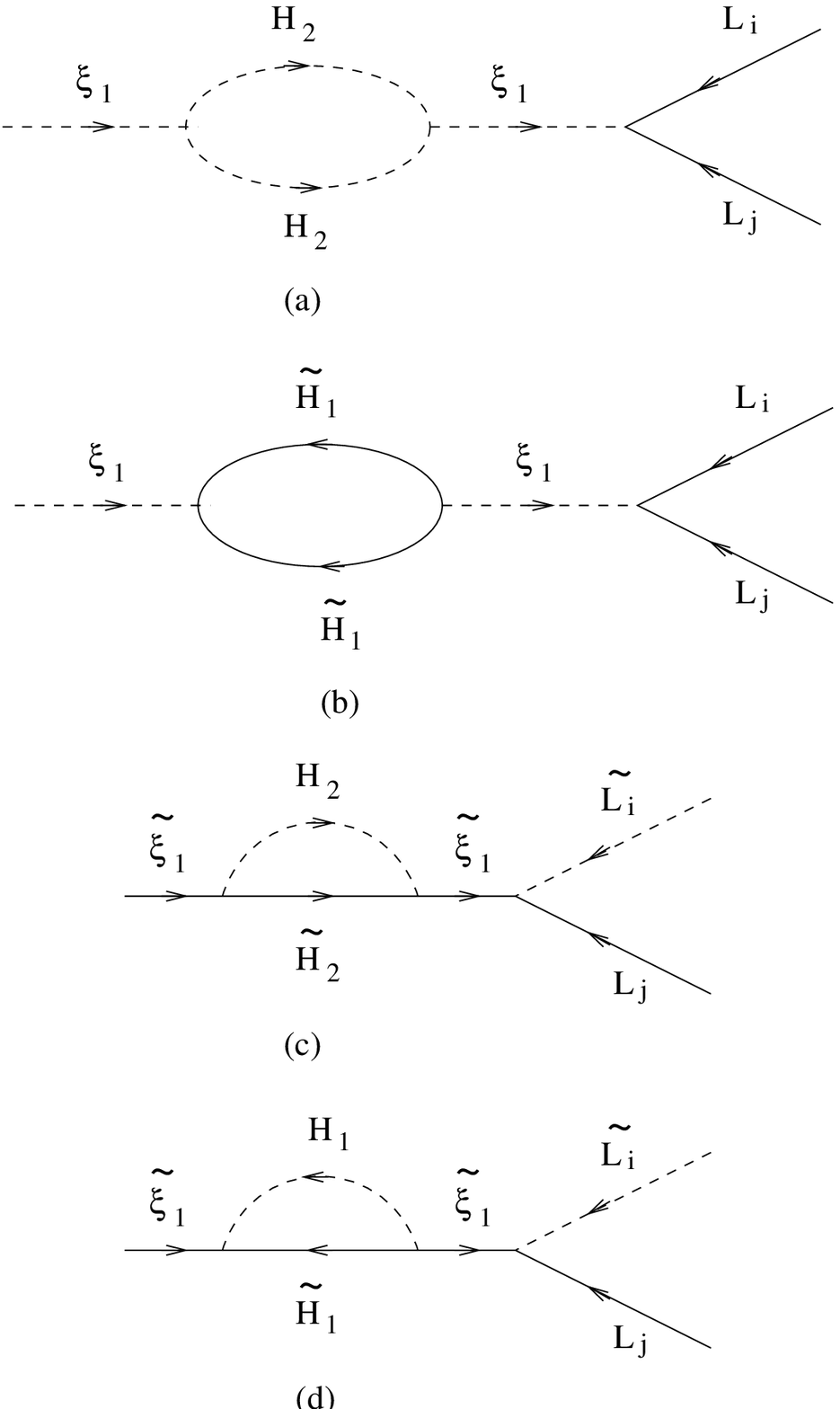}
}
\caption{ One loop diagrams contributing to CP violation in decays of 
$\hat \xi_1$. }
\label{strfg2}
\end{figure}

\begin{figure}[t]
\centerline{
\epsfxsize = 0.7\textwidth \epsffile{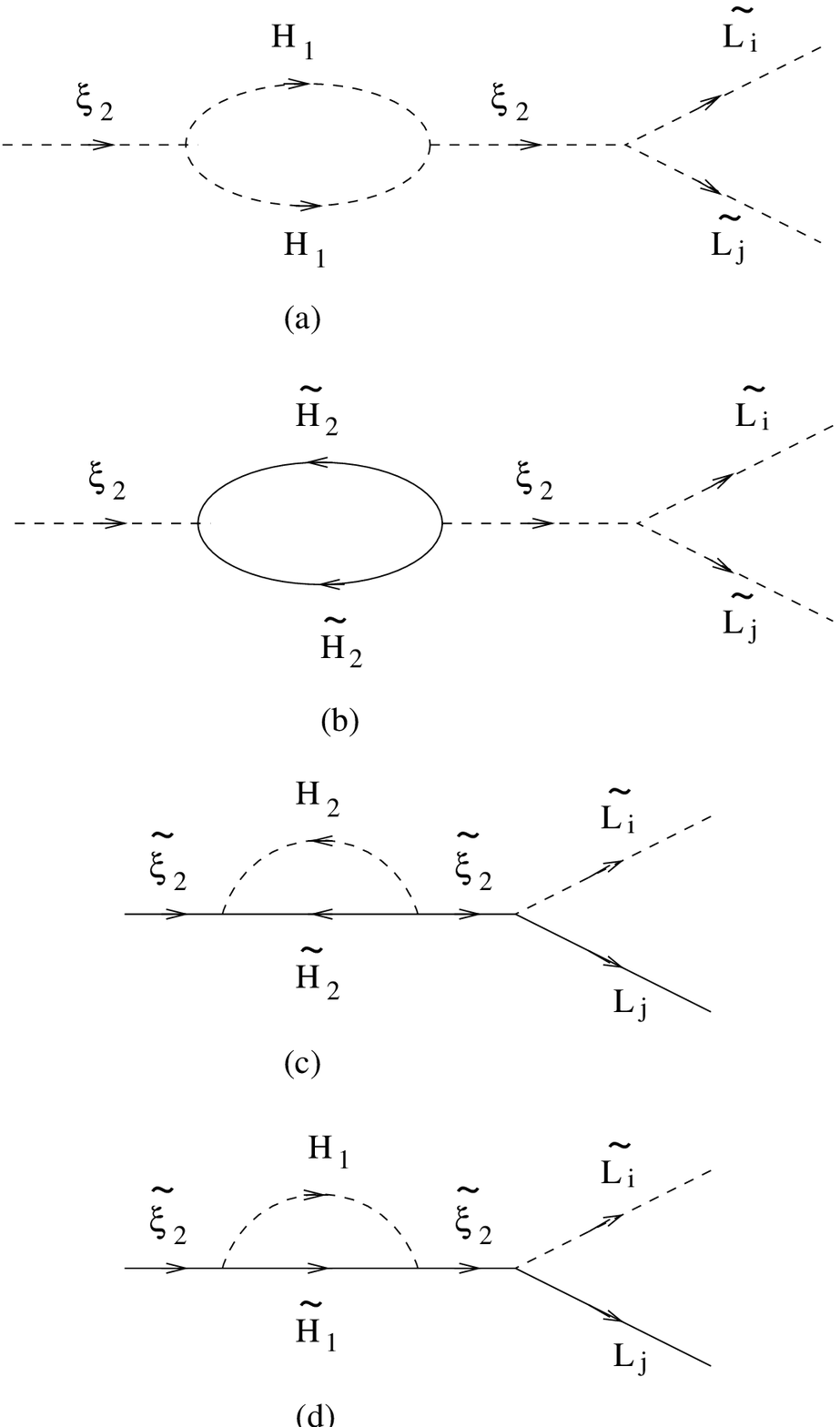}
}
\caption{One loop diagrams contributing to CP violation in decays of 
$\hat \xi_2$.  }
\label{strfg4}
\end{figure}

In the present case there are one-loop diagrams which
are given in Figure \ref{strfg2} (for $\xi_{1}^a$ and 
$\tilde \xi_{1}^a$ decays) and in Figure \ref{strfg4} (for 
$\xi_{2}^a$ and $\tilde \xi_{2}^a$ decays). 
As in the nonsupersymmetric case,
although some of the tree-level diagrams appear similar to the right-handed 
neutrino decay diagrams \cite{lepto}, there are no one-loop diagrams which are 
similar to the vertex diagrams of the right-handed neutrino decays. 
From this point of view, leptogenesis with the triplet Higgs 
scalars have this unique feature that CP violation comes only from 
the self-energy diagrams, which has the interpretation of
oscillations of the scalars before they decay \cite{paschos}.
Moreover, it was pointed out that the CP violation coming from the
self-energy diagrams has an interesting feature of resonant
oscillation.  Thus the amount of lepton asymmetry can get highly
enhanced when the masses of the triplet Higgs superfields are almost 
degenerate \cite{paschos}. 

In none of the loop diagrams of Fig.~3-4 is there any interference between
$\xi_1^a$ and $\xi_2^a$.  So, with one each of $\xi_1^a$ and $\xi_2^a$, 
there cannot be any CP violation. In this case, the relative phases
between various couplings can be chosen to be real. Only when there
are at least two $\xi_1^a$ or $\xi_2^a$, there can be CP violation. 
In this case, decays of both $\xi_1^a$ and $\xi_2^a$
will contribute to the amount of CP violation. The relative phases
between the couplings of the $\xi_1^a$ to the leptons of different
generations cannot generate a lepton asymmetry of the Universe, 
because they all correspond to final states of the same lepton number.
Among the loop
diagrams, Figures (c) and (d) are supersymmetric counterparts
of Figures (a) and (b), so supersymmetry ensures that the 
contributions from the first two diagrams are the same as that of
the last two diagrams. In the following we will consider 
explicitly only the decays of the scalar triplets keeping in
mind that the decays of their fermionic superpartners give the same
lepton asymmetry.

We shall now calculate the amount of CP violation generated
from the interference of the tree-level processes and the
one-loop diagrams. 
In the mass-matrix formalism, it is possible to give a 
physical interpretation to this CP violation. A triplet scalar superfield
oscillating into another type before it decays, has a different
decay rate compared to its conjugate states. 
Although the total decay
rates are equal by CPT, the partial decay rates now differ, which
give rise to CP violation.  This CP violation will then 
lead to a lepton asymmetry 
due to the fact that (1) the partial decay products
do not all have the same lepton number and (2) 
the interaction rate is not much faster than the
expansion rate of the Universe.

Without loss of generality, we shall assume that the mass matrix
for the triplet Higgs scalars starts out as real and diagonal, 
$$M_{ab}= M_a \delta_{ab} , $$ with $M_a$ real. However, in the
presence of interactions, they will no longer remain real. 
Including the interactions, the mass matrix for the left and 
right chiral superfields gets different contributions from the 
interference of the tree and loop diagrams. The physical states
of the left and right chiral superfields will evolve in a
different way and their decays into leptons and antileptons
would generate the lepton asymmetry of the Universe. 
In the following we denote by $\hat \phi^m_{1 +}$  and $\hat \phi^m_{2 +}$ 
with $m=1,2$ the physical states which are combinations
of the left chiral superfields $\hat \xi_1^a$ and $\hat \xi_2^a$ respectively,
and by $\hat \phi^m_{1 -}$ and $\hat \phi^m_{2 -}$ the physical states which 
are combinations of the conjugates of these 
superfields $\hat \xi_1^{a \ast}$ and $\hat \xi_2^{a \ast}$ respectively
(which are the right chiral superfields).

The effective scalar triplet mass matrix we obtain at one loop is given by 
\begin{equation}
\xi_1^{a \dagger} ({\cal M}^2_{1})_{ab} \xi_1^b +
\xi_2^{a \dagger} ({\cal M}^2_{2})_{ab} \xi_2^b 
\end{equation}
where, for a given value of the squared momemtum $p^2_\xi$ of the 
incoming or outgoing particle:
\begin{equation}
{\mathcal{M}}_{k }^2=\left(
\begin{array}{cc}
M^2_1 - i \Gamma^k_{11} M_1& 
 -i \Gamma^{k}_{12} M_2
 \\
-i \Gamma^{k}_{21} M_1 &
M_2^2 - i \Gamma^k_{22} M_2
\end{array} \right)\,,\label{massmatrix}
\end{equation}
with $\Gamma^{k }_{ab} M_b = (\Gamma^{k }_{ba})^* M_a$ and
$$ \Gamma^1_{ab} M_b = {1 \over 8 \pi} \left( 
\sum_{i,j} {f^a_{ij}}^* {f^b_{ij}} p^2_\xi +
 h_1^{a \ast} h_1^b p^2_\xi
+ M_a M_b h_2^{a} {h_2^{b \ast}}  
\right) , $$
and  
$$ \Gamma^2_{ab} M_b = {1 \over 8 \pi} \left( 
M_a M_b \sum_{i,j} f^a_{ij} f^{b \ast}_{ij} + 
h_2^{a \ast} h_2^b p^2_\xi
+  M_a M_b h_1^a h_1^{b \ast} 
\right) . $$
The decay widths $\Gamma_{\phi_{k \pm}^a}$  of the tree scalars in
the triplet $\phi_{k \pm}^a$ are given 
by $\Gamma_{\phi_{k \pm}^a}= \Gamma_k^{aa} \equiv \Gamma_{\phi_k^a}$.
Neglecting terms of order $[\Gamma_{ij} M_j / (M_1^2 - M_2^2)]^2$ 
the two mass matrices have the eigenvalues $M_{\phi_{k\pm}^a}=M_a$ 
and the eigenvectors are 
\begin{eqnarray}
\phi^1_{k +}&=& \xi_k^1 - i \frac{\Gamma^k_{12} M_2}{M_1^2 - M_2^2}
\xi_k^2 \label{mix1b}\\
\phi^2_{k +}&=&   i \frac{\Gamma^{k \ast}_{12} M_2}{M_1^2 - M_2^2}
\xi_k^1 +  \xi_k^2   \\
\phi^1_{k -}&=& \xi_k^{1 \ast} - i \frac{\Gamma^{k^ \ast}_{12} M_2}
{M_1^2 - M_2^2} \xi_k^{2 \ast}
  \\
\phi^2_{k -} &=&   i \frac{\Gamma^k_{12} M_2}{M_1^2 - M_2^2}
 \xi_k^{1 \ast}+  \xi_k^{2 \ast} . \label{mix4b}
\end{eqnarray}
Similarly we have
\begin{eqnarray}
\xi_k^1&=& \phi^1_{k +} + i \frac{\Gamma^k_{12} M_2}{M_1^2 - M_2^2}
\phi^2_{k +} \label{mix1}\\
\xi_k^2&=&   -i \frac{\Gamma^{k \ast}_{12} M_2}{M_1^2 - M_2^2}
\phi^1_{k +}  +  \phi^2_{k +}   \\
\xi_k^{1 \ast}&=&  \phi^1_{k -} + 
i \frac{\Gamma^{k  \ast}_{12} M_2}{M_1^2 - M_2^2}
\phi^2_{k -}   \\
\xi_k^{2 \ast}&=&   -i \frac{\Gamma^k_{12} M_2}{M_1^2 - M_2^2}
\phi^1_{k -}  +  \phi^2_{k -} . \label{mix4}
\end{eqnarray}
Note that, due to CP violation, the $\phi_{k-}^i$ are not  
Hermitian conjugates of the $\phi_{k+}^i$ but 
the orthonormality relations $\langle \phi_{k+}^i | \phi_{k-}^j 
\rangle =
\langle \phi_{k-}^i | \phi_{k+}^j \rangle =\delta_{ij}$ between the 
in and out states are satisfied
(as they should be) when diagonalizing a non-Hermitian 
mass matrix (see e.g. Refs.~\cite{sachs,beuthe}). 
The resulting lepton asymmetries $\varepsilon_k^m$ induced by the decay
of the scalar triplet $\phi_{k \pm}^a$ are given by
\begin{eqnarray}
\varepsilon_1^a &=& 2 \,   \frac{ \Gamma \left(\phi^a_{1 -} 
\to l l \right) - \Gamma \left(\phi^a_{1 +} \to l^c l^c \right)} 
{ \Gamma_{\phi^a_{1 -}} 
+ \Gamma_{\phi^a_{1 +}} },\\
\varepsilon_2^a &=&  2 \,  \frac{ \Gamma \left(\phi^a_{2 +} 
\to l l \right) - \Gamma \left(\phi^a_{2 -} \to l^c l^c \right)} 
{  \Gamma_{\phi_{2+}^a} 
 + \Gamma_{\phi^a_{2 -}} }.
\end{eqnarray}  

Putting Eqs.~(\ref{mix1})-(\ref{mix4}) in Eqs.~(\ref{superpot}) 
and (\ref{scalpot}) we obtain 
\begin{equation}
\varepsilon_1^a \simeq {1 \over 2 \pi (M_1^2 - M_2^2)}
\frac{ \sum_{i,j} \left\{  M^2_a  \, {\rm Im} [ h_1^2 {h_1^1}^* 
f^1_{ij} f^{2 \ast}_{ij} ] +
{\rm Im} [ M_2 M_1 h_2^{2 \ast} {h_2^1} 
f^1_{ij} f^{2 \ast}_{ij} ] \right\} }
{ 
\sum_{i,j} |f^a_{ij}|^2 +
|h_1^a|^2 +
|h_2^a|^2 
} \label{eps1}
\end{equation}
and similarly we have
\begin{equation}
\varepsilon_2^a \simeq {1 \over 2 \pi (M_2^2 - M_1^2)} 
{M_1 M_2 \over M^2_a} 
\frac{ \displaystyle{\sum_{i,j}} \left\{ 
M^2_a \,
{\rm Im} [ h_2^2 h_2^{1 \ast} 
f^{1 \ast}_{ij} f^{2}_{ij}] +
{\rm Im} [  M_1 M_2 h_1^{2 \ast} {h_1^{1}} 
f^{1 \ast}_{ij} f^{2}_{ij}] 
\right\} }
{ 
\sum_{i,j} | f^a_{ij} |^2 +
|h_2^a|^2 +
|h_1^a|^2 
} \label{eps2} 
\end{equation}
As expected the asymmetries come from the interference of the leptonic 
sector (through the $f_{ij}^a$'s)
and the non-leptonic sector (through the $h_k^a$'s).
Such asymmetries are obtained from the decay 
of each one of the tree states in each scalar triplet. 
Equal asymmetries are also obtained from the decay of the tree fermionic 
partners of the scalar triplets. Note that for $M_1$ close 
to $M_2$, $\varepsilon_1^a \sim \varepsilon_2^a$.  

When the mass difference between the two Higgs scalars is
very small and is comparable to the decay width, there is a
resonance in the amount of CP asymmetry, hence in the amount of 
lepton asymmetry. Our present
method fails in the limit when the decay width is larger than
the mass differences.
However as we did already in Eqs.~(\ref{mix1b})-(\ref{mix4}) we will restrict
ourselves to a region where the mass squared difference can be small
but still larger than the decay widths, so that the 
formalism we consider can be used safely.
Note that from earlier results in the calculation of
the resonance conditions, we understand that enhancement of the
asymmetry is almost maximal near the resonant 
condition $M_1^2-M_2^2 \sim \Gamma^k_{12} M_2$. So,
extending the analysis to even smaller mass difference would not 
improve our result in any case. 

\section{Boltzmann Equations}

We shall now check if the out-of-equilibrium condition is 
satisfied in this scenario and can generate the required amount
of baryon asymmetry of the Universe. The naive consideration
for the out-of-equilibrium condition that the decay rates of
the triplet Higgs scalars to be less than the expansion rate
of the universe is satisfied for a wide range of parameters.
This out-of-equilibrium condition reads,
\begin{equation}
K_{\phi^a_k} = {\Gamma_{\phi_k^a} \over H(M_a)} < 1  
\end{equation}
where the Hubble constant $H(T)$ at the temperature $T$ is given by
\begin{equation}
H(T)=\sqrt{\frac{4 \pi^3 g_\ast}{45}} \frac{T^2}{M_{P}},\label{hubble}
\end{equation}
with $g_\ast \sim 100$ the number of massless degrees of freedom
and $M_{P} \sim 10^{19}$ GeV is the Planck scale.  
Given any particular
temperature, the out-of-equilibrium condition constrains the various
coupling constants. 
If this condition is satisfied
and if the various damping terms 
due to scatterings are negligible, 
the total amount of lepton
asymmetry per comoving volume $X_L \equiv n_L/s = (n_l - n_{\bar{l}} )/s$ 
that will be generated through the decays of the four
triplet Higgs superfields will be given by 
$\sum_k 6 (\varepsilon_1^k + \varepsilon_2^k) n_\gamma/(2s)= 
\sum_k 6 (\varepsilon_1^k+\varepsilon_2^k) 45/(2 g_\ast \pi^4)$
where the entropy $s$ and the photon number density $n_\gamma$
are given by
\begin{eqnarray}
s &=& g_* {2 \pi^2 \over 45} T^3 , \\
n_\gamma&=&\frac{2 \, T^3}{\pi^2}.
\end{eqnarray}
For the out-of-equilibrium condition of $\phi_{1,2}^a$ to be satisfied,
we get a bound on the parameters
\begin{equation}
\frac{ \sum_{i,j} |f^a_{ij}|^2 + |h_1^a|^2 + |h_2^a|^2 }{M_a} < 
\sqrt{\frac{4 \pi^3 g_\ast}
{45}} \frac{8 \pi}{M_P} \sim (4 \cdot 10^{-17} \,\, 
\mbox{GeV}^{-1} ) .
\label{ooec}
\end{equation}
It is interesting to compare this condition with the condition 
that a neutrino mass of order $\sim 10^{-3}$~eV
is generated from Eq.~(\ref{numass}),
\begin{equation}
- \sum_a \frac{f_{ij}^a h^a_2}{M_a} = \frac{(m_{\nu})_{ij}}{2 v_2^2}
\sim (10^{-17} \,\, \mbox{GeV}^{-1} ),
\label{numasscond}
\end{equation} 
where $v_2$ has to be of order $v=246$~GeV.
A neutrino mass of order $10^{-3}$ eV can therefore be obtained while
the out-of-equilibrium condition is satisfied for any value of $M_1$ and $M_2$ 
provided the couplings $f^a_{i,j}$ and $h^a_{1,2}$ have the 
appropriate values\footnote{For small values of $M_{1,2}$ this 
would require however very small values of the $f$'s and the $h$'s whose 
naturalness could be questioned.}.
This is in general achieved if $h^a_2$ together 
with at least one of 
the $f^a_{ij}$ for $a=1$ or 2 are of 
order $\sim [ (10^{-17} \cdot \, \mbox{GeV}^{-1}) M_{1,2} ]^{1/2}$ (with 
all other couplings taking smaller values).
For $M_{1,2} \sim 10^{14}$ GeV this requires 
$f_{ij}^a \sim h_2^a \sim 10^{-2}$-$10^{-1}$ while for
$M_{1,2} \sim 10^{9}$ GeV this requires 
$f_{ij}^a \sim h_2^a \sim 10^{-3}$-$10^{-4}$.
Assuming a maximal CP violating phase, 
the lepton asymmetry obtained from Eqs. (\ref{eps1})-(\ref{eps2})
is then typically of order $X_L \sim 10^{-5}$-$10^{-6}$ in the former case
and $X_L \sim 10^{-10}$-$10^{-11}$ in the latter case. 
A smaller asymmetry can be generated if for example this CP violating phase 
is not maximal or if
in general larger values of the $f$'s and the $h$'s are taken in such a way 
that Eq.~(\ref{numasscond}) is satisfied but not Eq.~(\ref{ooec}). In the 
latter case the damping term of the inverse decay process will
suppress the asymmetry. A larger asymmetry can be obtained if $M_1$ and $M_2$
are more degenerate. For $M_{1,2} < 10^{9}$ GeV a certain degree of degeneracy
is needed in order to obtain a baryon asymmetry of the order of the 
one required, i.e. $X_L \sim 10^{-10}$.

The above estimate has however not taken into account possible 
scattering damping terms.
There are for example 
lepton-number violating scattering processes, which
can deplete the generated lepton asymmetry of the
Universe. For example, $H_1 + H_1 \to \tilde L + \tilde L$
and $H_1 + H_1 \to \xi_2 \to \tilde L + \tilde L$ scattering (which 
are absent in the nonsupersymmetric case) come from 
renormalizable terms and may not be suppressed.   However, it can be shown 
that these processes are not really relevant because they go out of 
equilibrium once we require the decays of the triplets to be slow enough to 
be away from thermal equilibrium.  There is also one lepton-number 
conserving process which is more of a problem, i.e. the gauge interactions 
of the triplet Higgs superfields.  These induce the very fast
$\xi_a^\dagger + \xi'_a \to G_1 + G_2$ scattering process, 
where $\xi_a$ and $\xi'_a$ are two scalar triplets, and $G_1$ and $G_2$ are 
two $SU(2)_L$ or $U(1)_Y$ gauge bosons, as obtained from the kinetic term
of the scalar triplets.
This gives a suppression in the generation
of the lepton asymmetry of the Universe and implies that the mass 
of the triplets cannot be too small (except if the two triplets are
almost degenerate as shown below). The presence of this damping term
requires the explicit calculation of the evolution of the asymmetry using
the Boltzmann equations.
 
Defining the variable $z \equiv M_1/T$ and 
the various number densities per comoving volume $X_i \equiv n_i/s$, 
the Boltzmann equations are:
\begin{eqnarray}
\frac{dX_{{\phi}_k^a}}{d z}&=& 
- z K_{{\phi}_k^a} \frac{K_1(z)}{K_2(z)}   
\Big( \frac{M_{{\phi}_k^a}}{M_1}  \Big)^2 
\Big({X_{{\phi}_k^a}}-{X^{eq}_{{\phi}_k^a}}  \Big)
+ z  \frac{1}{s H(M_1)} \Big( 1- \frac{X_{{\phi}_k^a}^2}
{X_{{\phi}_k^a}^{eq2}}  \Big)
\gamma^{a}_{scatt.} 
\label{boltzmann1}\\
\frac{d X_L}{d z}&=& \sum_{a,k}
z K_{{\phi}_k^a}   \frac{K_1(z)}{K_2(z)} 
\Big( \frac{M_{{\phi}_k^a}}{M_1}  \Big)^2 
\Big[\varepsilon_k^a ({X_{{\phi}_k^a}}-{X^{eq}_{{\phi}_k^a}} )
- \frac{1}{2}   \frac{X^{eq}_{{\phi}_k^a}}{X_\gamma} X_L \Big]
.\label{boltzmann2}
\end{eqnarray} 
In Eqs.~(\ref{boltzmann1})-(\ref{boltzmann2})
the equilibrium distributions of the number 
densities are given by the Maxwell-Boltzmann statistics:
\begin{equation}
n_{{\phi}_k^a}=g_{{\phi}_k^a}\frac{M_{{\phi}_k^a}^2}{2 \pi^2}
T K_2(M_{{\phi}_k^a}/T), 
\end{equation}
where $g_{{\phi}_k^a}=1$ are the 
numbers of degrees of freedom of the ${{\phi}_k^a}$ and 
$K_{1,2}$ are the usual modified Bessel functions.  The reaction 
density for the scattering process $\xi_a^\dagger + \xi'_a \to G_1 + G_2$
is given by
\begin{equation}
\gamma^{a}_{scatt.}= \frac{T}{64 \pi^4} \int_{4 M_a^2}^{\infty}
ds\, \hat{\sigma}_a(s)\, \sqrt{s} \, K_1(\sqrt{s}/T),
\label{scattint}
\end{equation}
where $\hat{\sigma}$ is the reduced cross section which is given 
by $2 ( s - 4 M_a^2 ) \sigma_a (s)$.
Note that a precise result would require an explicit calculation of
all scattering processes involving gauge interactions in all 
channels.\footnote{There 
are more than 20 different physical processes of the type
$\xi_a^\dagger + \xi'_a \to G_1 + G_2$. There is also scattering of the 
type $\xi_a^\dagger + \xi'_a \rightarrow l + \bar{l}$
with an intermediate gauge boson which is of the same order.} 
However it can be checked that the dependence of the generated lepton 
asymmetry on the magnitude of the scattering is much slower than linear. 
Therefore, considering also the fact that the model allows some freedom in 
the range of parameters used, this explicit calculation will not add much 
to our understanding in any case.  We will thus make the following estimate: 
\begin{equation}
\sigma_a = \frac{1}{\pi \sqrt{s}} \frac{1}{\sqrt{s-4 M_a^2}} \, g^4
\label{scattres}
\end{equation}
where $g$ is the $SU(2)_L$ coupling (which at tree level is given
by the relation $m_W^2=g^2 v^2/4$). 
Putting Eq.~(\ref{scattres}) in Eqs.~(\ref{scattint}) and (\ref{boltzmann1}), 
it turns out that the scattering term has a small effect on the evolution of 
the lepton asymmetry for values 
of $M_{1,2}$ above $10^{11}-10^{12}$~GeV. For smaller values of $M_{1,2}$
the suppression can be very strong due to the fact that the last term 
of Eq.~(\ref{boltzmann1}) increases when $M_1$ decreases and $T \sim M_1$.
This will suppress the asymmetry which at some point becomes much smaller than 
$\sim 10^{-10}$ except if $M_1$ and $M_2$ are sufficiently degenerate. 
Note that the suppression due to these scattering processes is the most 
effective when the triplet starts decaying.  At lower temperatures, the 
scattering effect is suppressed by the Boltzmann factor due to the higher 
threshold in Eq.~(\ref{scattint}).  Taking for example $f_{22}^a$, $f_{23}^a$, 
$f_{32}^a$ and $f_{33}^a$ (for both $a=1$ and 2) equal to the same value $f$ 
with all other $f_{ij}^a$ equal to zero, i.e. assuming negligible 
all the $f^a_{ij}$ with $i=1$ and/or $j=1$, (which constitutes one of the 
possible structures leading to a maximal mixing between the second and third 
generation of neutrinos) and taking all $h_k^a$ couplings 
equal to the same value $h$, 
four typical sets of parameters which give an asymmetry of order  
$\sim 10^{-10}$ together with a neutrino mass of order 
$10^{-3}-10^{-2}$~eV are shown below:
\begin{eqnarray}
M_1&=&10^{13}\,\, \mbox{GeV} \,\,\, 
M_2= 3.0 \cdot 10^{13} \,\, \mbox{GeV} \,\,\, h=1 \cdot 10^{-3} \,\,\,
f=3 \cdot 10^{-2}\\
M_1&=&10^{12}\,\, \mbox{GeV} \,\,\, 
M_2= 3.0 \cdot 10^{12} \,\, \mbox{GeV} \,\,\, h=1 \cdot 10^{-2} \,\,\,
f=5 \cdot 10^{-4}\\
M_1&=&10^{11}\,\, \mbox{GeV} \,\,\, 
M_2= 2.0 \cdot 10^{11} \,\, \mbox{GeV} \,\,\, h=8 \cdot 10^{-4} \,\,\,
f=2 \cdot 10^{-3}\\
M_1&=&10^{10}\,\, \mbox{GeV} \,\,\, M_2= 1.1 \cdot 10^{10}\,\, \mbox{GeV}
\,\,\, h=1 \cdot 10^{-3} \,\,\, f=5 \cdot 10^{-4}. \label{set4}
\end{eqnarray}
A maximal CP-violating phase has been assumed.
Note that the 
degree of degeneracy which is required for $M_{1,2} \sim 10^{10}$~GeV 
is relatively small. Note also that smaller values 
of $M_{1,2}$ are possible if they are even more degenerate.  As 
$M_{1,2}$ decreases, the degree of degeneracy required becomes 
however very high, due to the damping effects of the scattering processes.

\section{Gravitino Problem}

So far we have not taken into account the gravitino problem. The main 
constraint comes from the fact that the lepton asymmetry has to be generated 
after inflation, which is very important in supersymmetric models
\cite{gravitino,rh1}. 
The thermal production of massive gravitinos restricts the 
beginning of the radiation-dominated era following inflation. 
The reheating temperature after inflation
is constrained by requiring gravitino production to be suppressed so that 
it will not overpopulate the Universe.  Since the 
gravitinos interact very weakly, they decay very late and modify
the abundances of light elements which may become inconsistent
with nucleosynthesis.  On the other hand, if they are stable, then
they overclose the universe. The upper bound on the reheating 
temperature from the gravitino constraint is \cite{rh1}
\begin{equation}
T_{RH} \leq 10^{10} ~{\rm GeV}~ \times \left({ m_{3/2} \over 
100 ~{\rm GeV} } \right) \times \left( {1 ~{\rm TeV} \over
m_{\tilde g}(\mu) } \right)^2
\end{equation}
where $m_{3/2}$ is the gravitino mass and $m_{\tilde g}$ is the
running mass of the gluino. This gravitino constraint is satisfied
if the lepton asymmetry is generated at temperatures below the
reheating temperature
$T < T_{RH}$. 
From this result we can assume that the masses of the triplet Higgs scalars 
should be around $T_{RH} \sim 10^{10}-10^{11}$ GeV, 
so that leptogenesis occurs at a temperature $T < 10^{10}-10^{11}$ GeV.
As shown above, a lepton asymmetry and neutrino masses of the size 
required can be generated with this value of the mass, but it requires some 
(moderate) degree of degeneracy between $M_1$ and $M_2$ [see Eq.~(\ref{set4})].

\section{Summary and Conclusion}

We have shown that the supersymmetric triplet Higgs model developed in this 
article constitutes an interesting and simple alternative for generating 
neutrino masses and baryogenesis. This requires typically triplet superfields 
with mass of order $10^{9}-10^{14}$~GeV. We have shown that this mechanism 
has the interesting property of possible resonant behavior. For masses of 
order $10^{10}$~GeV which are consistent with the gravitino 
problem, this feature of resonant CP violation is necessary (i.e. masses 
of the triplets need to be moderatly degenerate).

\vskip 0.5in
\begin{center} {ACKNOWLEDGEMENT}
\end{center}

This work was supported in part by the U.~S.~Department 
of Energy under Grant No.~DE-FG03-94ER40837 and by the 
TMR, EC-contract No.~ERBFMRX-CT980169(Euro\-Da$\phi$ne). U.S. and T.H. thank
the Physics Department, University of California at Riverside 
for hospitality.

\newpage
\bibliographystyle{unsrt}

\end{document}